\documentclass[sigconf]{acmart}
\usepackage{bm}
\usepackage[whole]{bxcjkjatype} 

\usepackage{multirow}
\usepackage{tabularx}
\usepackage{hyperref}

\AtBeginDocument{%
  }

\setcopyright{acmlicensed}
\copyrightyear{2024}
\acmYear{2024}
\acmDOI{XXXXXXX.XXXXXXX}

\begin{document}

\title{Multimodal Point-of-Interest Recommendation}

\author{Yuta Kanzawa}
\email{tyutakanzawa@gmail.com}
\affiliation{%
  \institution{Independent Consultant}
\city{Tokyo}
\country{Japan}
}

\author{Toyotaro Suzumura}
\affiliation{%
  \institution{The University of Tokyo}
  \streetaddress{}
  \city{Tokyo}
  \country{Japan}}
\email{suzumurat@gmail.com}

\author{Hiroki Kanezashi}
\affiliation{%
  \institution{The University of Tokyo}
  \streetaddress{}
  \city{Tokyo}
  \country{Japan}}
\email{hkanezashi@ds.itc.u-tokyo.ac.jp}

\author{Jiawei Yong}
\affiliation{%
  \institution{Toyota Motor Corporation}
  \streetaddress{}
  \city{Tokyo}
  \country{Japan}}
\email{jiawei_yong@mail.toyota.co.jp}

\author{Shintaro Fukushima}
\affiliation{%
  \institution{Toyota Motor Corporation}
  \streetaddress{}
  \city{Tokyo}
  \country{Japan}}
\email{s_fukushima@mail.toyota.co.jp}


\begin{abstract}
  Large Language Models are applied to recommendation tasks such as items to buy and news articles to read. Point of Interest is quite a new area to sequential recommendation based on language representations of multimodal datasets. As a first step to prove our concepts, we focused on restaurant recommendation based on each user's past visit history. When choosing a next restaurant to visit, a user would consider genre and location of the venue and, if available, pictures of dishes served there. We created a pseudo restaurant check-in history dataset from the Foursquare dataset and the FoodX-251 dataset by converting pictures into text descriptions with a multimodal model called LLaVA, and used a language-based sequential recommendation framework named Recformer proposed in 2023. A model trained on this semi-multimodal dataset has outperformed another model trained on the same dataset without picture descriptions. This suggests that this semi-multimodal model reflects actual human behaviours and that our path to a multimodal recommendation model is in the right direction.
\end{abstract}




\maketitle

\section{Introduction}
Recommendation systems have been developing recently and widely used in web services such as e-commerce and online news \cite{zhou2023comprehensive-survey}. Current prominence of Large Language Models (LLMs) leads to their use in sequential recommendation tasks where traditional recommendation systems have been dominant. OpenP5 \cite{xu2023openp5} learns a sequence of item IDs as a sequence of words and generates the next item ID. Recformer \cite{li2023text-is-all-you-need} learns a sequence of item attributes such as name and colour as a language representation, and it can be applied to multimodal sequential recommendations.

We aim to develop a model to address various challenges in the mobility sector, especially Point of Interest (POI) recommendations. The sector is inherently multimodal with abundant data types such as trajectory data, audio and visual data while it is quite a new area to sequential recommendation based on multimodal datasets. Many models have been proposed but most of them are based on POI IDs \cite{islam2020survey, yin2023next-poi-recommendation}. We assume visual information would increase model performance in POI recommendation. Besides, POI recommendation has geographical constraints unlike the item recommendation in e-commerce. A good POI recommendation model should propose a venue in a user's local area rather than a most suitable venue to this user far away. Spatial index of venues could address it.

As a first step to achieve our POI recommendation model based on textual information as well as visual and geographic attributes, we focus on restaurant recommendation based on each user's past visit history and venues' location and food descriptions. A user would think about genre and location for the next meal and check what kind of dishes are served if comments or pictures of them are available. We mock this process by training a sequential recommendation LLM (Recformer) with restaurant attributes (name, genre, location) and textual descriptions of food images. We train 2 Recformer models with and without this pseudo-descriptions and the model trained with the food descriptions shows much better performance. In general, the contributions of this work can be listed as follows:

\begin{itemize}
    \item We propose a new application of sequential multimodal recommendation to POI sector.
    \item We introduce a geographical key of POI to address spatial constraints in POI recommendation tasks.
    \item We present experiments on a dataset demonstrating the performance of the proposed method.
\end{itemize}

\section{Data Analysis and Motivation}
To evaluate our concepts, we need a dataset of POI sequences with attributes presented in natural language. We choose the Foursquare dataset \cite{yang2015modeling-user-activity-preference} and the FoodX-251 dataset \cite{kaur2019foodx-251}. We use the Foursquare dataset to extract actual sequences of restaurant visits by users and the FoodX-251 dataset to mock descriptions and comments about foods served there.
\subsection{Dataset Preparation}
\label{dataset-preparation}
\begin{figure}[h]
  \centering
  \includegraphics[width=0.8\linewidth]{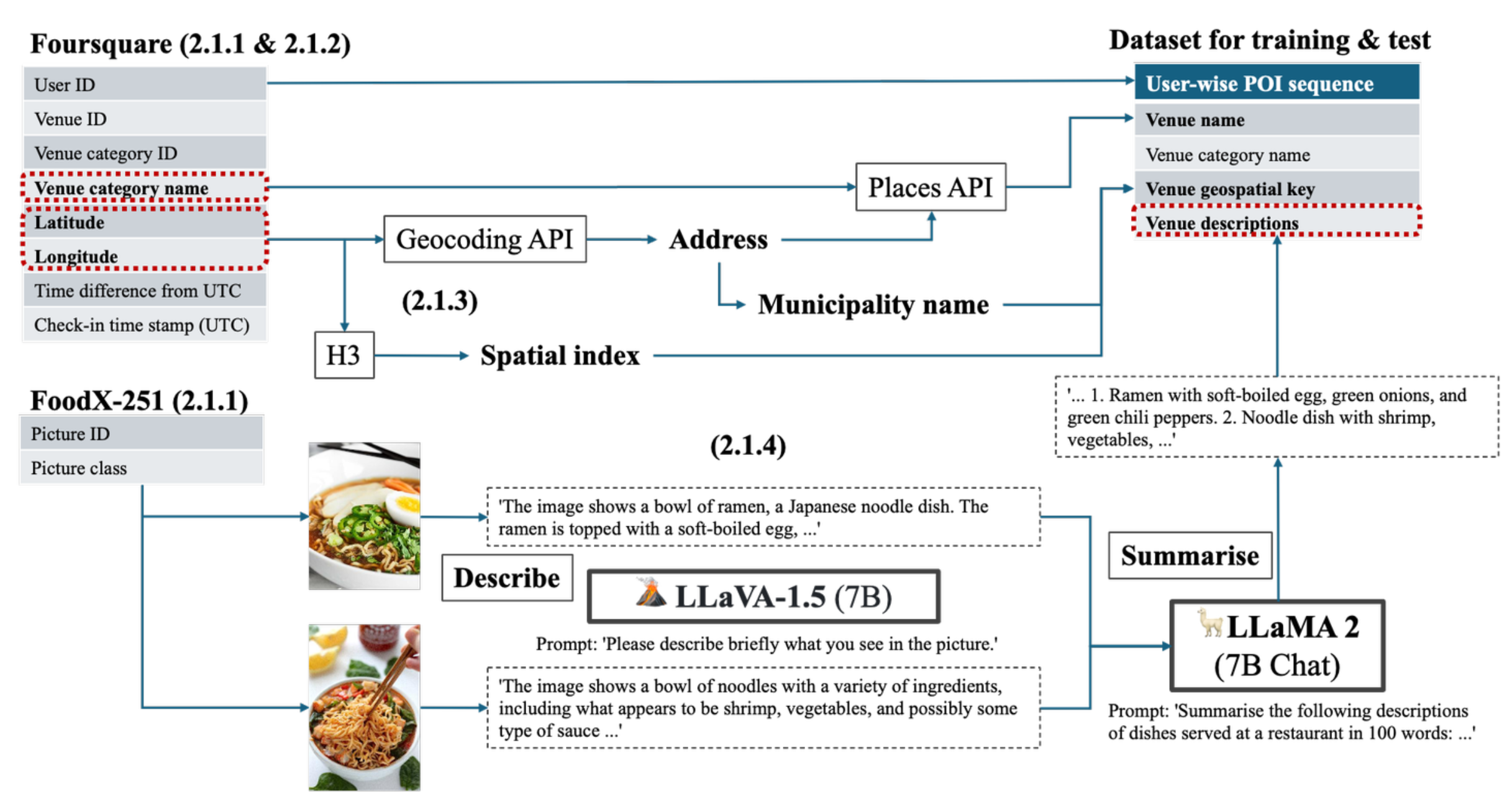}
  \caption{Dataset preparation}
  \label{fig:dataset_preparation}
\end{figure}

Figure \ref{fig:dataset_preparation} shows the whole process to prepare the dataset. We first select base datasets (\ref{base-datasets}), extract check-ins to food-related POIs (\ref{meta-data-and-interaction-data}), add geographical information to venues by using APIs (\ref{geographical-information}), and also generate pseudo-food descriptions of the venues with LLMs (\ref{food-descriptions}). The mapping data, including the venue IDs and addresses (latitude and longitude) of restaurants listed in the Foursquare dataset, the corresponding image files from FoodX-251 dataset, and description texts of the restaurants including images, is publicly available \footnote{\url{https://github.com/toyolabo/multimodal-poi}}.

\subsubsection{Base Datasets}
\label{base-datasets}
We choose the Foursquare dataset and FoodX-251 dataset (Table \ref{table:datasets}). The Foursquare dataset contains check-ins in New York City, United States and Tokyo, Japan from 12 April 2012 to 16 February 2013, and we used check-ins in Tokyo (573,703 in total). Each check-in has a user ID, a time stamp, a set of geographical coordinates (latitude and longitude), a venue ID, a venue category (a pair of a name and an ID). The venue categories are very detailed and we select food-related ones manually (\ref{meta-data-and-interaction-data}).
The FoodX-251 dataset contains 158,846 food images collected from the web and labelled in 251 classes. We select 106 classes based on check-in frequency in corresponding Foursquare's venue category (\ref{food-descriptions}).

\begin{table}[h]
    \caption{Description of base datasets}
    \label{table:datasets}
    \begin{tabularx}{\linewidth}{lXX}
        \hline
        \textbf{Name} & \textbf{Contents} & \textbf{Size of whole dataset}\\
        \hline
        Foursquare & User ID, venue ID, venue categories, geographical coordinates, check-in time stamps & 573,703 check-ins in Tokyo\\
        \hline
        FoodX-251 & Food images collected from the web & 158,846 images\\
        \hline
    \end{tabularx}
\end{table}

\subsubsection{Meta Data and Interaction Data}
\label{meta-data-and-interaction-data}
To extract check-ins at food-related POIs, we select 80 out of 247 categories defined in the Foursquare dataset (Appendix \ref{appendix-food-categories}), and there are 31,105 POIs under these categories. We finally get 28,989 POIs with attributes as explained later (we call this attribute dataset as meta data.) We extract check-ins at these POIs from the Foursquare dataset, filter check-ins by loyal users with 100 or more check-ins and get 119,105 check-ins by 2,092 users. 

\subsubsection{Geographic Information}
\label{geographical-information}
To generate a spatial index with language representation of geographical area of each POI, we first get an address of each POI by passing latitude and longitude to Google's Geocoding API \cite{google2023geocoding}, and then identify venue names of 28,989 POIs by passing an address and a venue category of each POI to Google's Places API \cite{google2023places}. The API also returned additional attributes and we keep one named 'types' which consists of short descriptions of each venue such as 'food', 'meal\_takeaway', 'establishment' for later use. Second, we convert postal codes contained in the addresses into municipality names (in Japanese) based on the official postal code list provided by Japan Post \cite{jp2023postalcodes}. Third, we get a spatial index of each POI from its latitude and longitude based on a system called H3 by Uber Technologies \cite{uber2024h3} which splits the globe into hexagonal areas.
Finally, we combine the municipality name and the spatial index, and get a geospatial key of each POI (e.g. '新宿区 882f5a3751fffff').

\subsubsection{Food Descriptions}
\label{food-descriptions}
To generate pseudo-food descriptions served at the venues, first, we map FoodX-251's image categories to Foursquare's venue categories allowing duplicates (up to 3). For example, we map 'hamburger' in FoodX-251 to 'American Restaurant', 'Fast Food Restaurant' and 'Burger Joint' in Foursquare. FoodX-251 has 251 categories and Foursquare has 80 categories, but both have less frequent or too specific or too wide categories. To achieve general and effective mapping between them, we selected 106 FoodX-251 categories and 30 Foursquare categories (Appendix \ref{appendix-4sq-foodx-mapping}).

Second, we randomly select the same numbers of food pictures as unique POIs in each venue category (at least 100) to populate the final meta data with unique food picture descriptions as much as possible while saving time of image-to-text conversion, and then randomly allocate 8 different pictures to each POI (a picture can be allocated to 2 or more POIs).

Third, we convert the allocated pictures into short descriptions using LLaVA (LLaVA-1.5 7B) \cite{liu2023llava} and LLaMA (LLaMA 2 7B Chat) \cite{touvron2023llama} in the following steps and build semi-multimodal dataset:
\begin{enumerate}
    \item Extract unique set of the allocated pictures to save time (10,741 pictures).
    \item Ask LLaVA to describe a picture with a prompt: 'Please describe briefly what you see in the picture.' Repeat this for each picture.
    \item Combine 8 descriptions for each POI.
    \item Ask LLaMA to generate a short food description from the combined descriptions with a prompt: 'Summarise the following descriptions of dishes served at a restaurant in 100 words:' Repeat this for each POI.
\end{enumerate}

Table \ref{table:descriptions} shows an example of the short descriptions. The models successfully describe and summarise food names and ingredients as well as their colours and styles.

\begin{table*}[h]
    \caption{An example of summarised description of pictures}
    \label{table:descriptions}
    \begin{tabularx}{\linewidth}{lX}
        \hline
        \textbf{POI category} & \textbf{Description}\\
        \hline
        Ramen / Noodle House & 'Here are the summaries of the dishes described in the images, in 100 words or less: 1. Ramen with soft-boiled egg, green onions, and green chili peppers. 2. Noodle dish with shrimp, vegetables, and sauce, eaten with chopsticks. 3. Ramen with soft-boiled egg, slices of bacon, and green leafy vegetable. 4. Soup with broccoli, mushrooms, and other vegetables, eaten with chopsticks. 5. Ramen with slices of meat and green onions, accompanied by chopsticks. 6. Rich, brown broth with slices of pork, soft-boiled egg, and green onions, eaten with chopsticks. 7. Creamy soup or sauce with noodles, eaten with a spoon. 8. Ramen with a variety of toppings, including green leafy vegetable.'\\
        \hline
    \end{tabularx}
\end{table*}

\subsection{Data Characteristics}
We have 111,801 check-ins by 2,092 users in the interaction data corresponding to 28,989 venues in the meta data. Average length of the food description is 686.7 characters (Table \ref{table:attributes}). Average and maximum check-in count is 250.2 and 2,991, and more than 75\% of users check in 300 or less times (N.B., we limit the minimum check-in frequency to 100 to extract loyal users).

\section{Methodology}

\subsection{Problem Formulation}
\label{problem-formulation}
In the datasets we prepare in \ref{dataset-preparation}, we have a POI set $P$ and a user's check-in sequence $s=\left\{p_1,p_2,...,p_n\right\}$ in the order of check-in time where $n$ is the length of sequence $s$ and $p$ is an element of set $P$. POI recommendation is a task to predict the next $p$ based on a sequence $s$. In the similar way of the original Recformer setting \cite{li2023text-is-all-you-need}, each POI $p$ has a corresponding dictionary of attributes $D_p$ which consists of key-value pairs $\left\{\left(k_1,v_1\right),\left(k_2,v_2\right),...,\left(k_m,v_m\right)\right\}$ where $k$ is an attribute name such as 'venue\_category', $v$ is the corresponding value such as 'French Restaurant' and $m$ is the number of attributes in the model. Both of $k$ and $v$ are in the form of natural languages (except the geospatial keys in 2.1.3) and consist of words $(k,v)=\left\{w_1^k,...,w_c^k,w_1^v,...,w_c^v\right\}$ where $w^k$ and $w^v$ are words in the vocabulary of Recformer and  is the length of a truncated attribute text $kv$ (e.g., 'venue\_categoryFrench Restaurant'). To input attributes in the dictionary $D_p$ into Recformer, we combine all the key-value pairs into a text $T_p=\left\{k_1,v_1,k_2,v_2,...,k_m,v_m\right\}$. The language model learns this language representation of attributes $T_p$ for each POI $p$ in a user's check-in sequence $s$ for each user.

\subsection{Recformer}
\label{recformer}
As explained in \ref{problem-formulation}, the model input is a sequence of truncated attribute texts of each venue. We conduct pretraining with Masked Language Modelling \cite{devlin2019bert-pre-training} and item-item contrastive task which are widely used to predict the next item to recommend, and then take Two-Stage Finetuning \cite{li2023text-is-all-you-need} with item-item contrastive learning as learning task. We keep the task settings of the previous work \cite{li2023text-is-all-you-need}.

\section{Experiments}
Our research question is adding visual information to model input would increase model performances in POI recommendation.

\subsection{Experimental Setup}
\begin{figure}[h]
  \centering
  \includegraphics[width=0.8\linewidth]{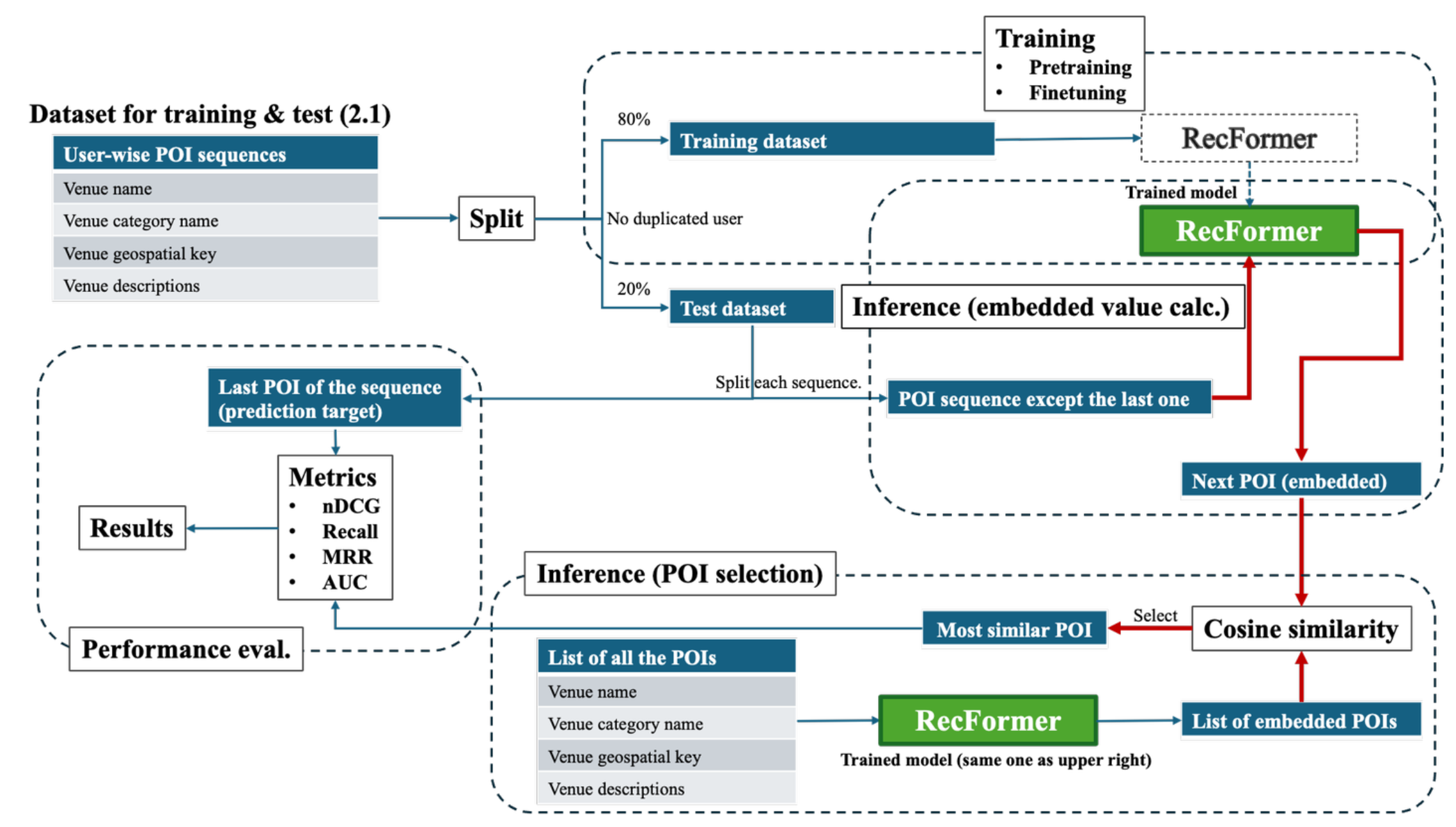}
  \caption{Experimental setup}
  \label{fig:experimental_setup}
\end{figure}

Figure \ref{fig:experimental_setup} shows the whole process. We take the same steps as the previous work \cite{li2023text-is-all-you-need}.
\begin{enumerate}
    \item Dataset split (\ref{dataset-split})
    \item Training (\ref{recformer}, \ref{training})
        \begin{enumerate}
            \item Pretraining
            \item Finetuning
        \end{enumerate}
    \item Inference (\ref{inference})
        \begin{enumerate}
            \item Prepare a list of embedded values of all the POIs.
            \label{inference-preparation}
            \item Take a sequence except the last POI.
            \item Calculate an embedded value of the next POI.
            \item Select the most similar POI from the list in \ref{inference-preparation}.
        \end{enumerate}
    \item Performance evaluation (\ref{performance-evaluation})
\end{enumerate}

\subsubsection{Dataset split}
\label{dataset-split}
We split the interaction data into a training dataset and a test dataset where the training dataset takes approximately 80\% of the original dataset. To avoid potential leakage, we make no duplicated user between the 2 datasets by the group shuffle split in the scikit-learn package\footnote{\url{https://scikit-learn.org/stable/modules/generated/sklearn.model_selection.GroupShuffleSplit.html}}.

\subsubsection{Training}
\label{training}
We train 2 Recformer models to evaluate model performances with and without picture descriptions (N.B., the 'types' from Places API are lists of keywords related to the venues and are included in both models' inputs.) Table \ref{table:attributes} shows the list of 4 attributes. The maximum number of tokens for each attribute is 1,024 and 2,048 for each check-in sequence.

\begin{table}[h!]
    \caption{Attributes}
    \label{table:attributes}
    \begin{tabularx}{\linewidth}{lXX}
        \hline
        \textbf{Name} & \textbf{Description} & \textbf{Average length (characters)}\\
        \hline
        venue\_category & Venue category in Foursquare & 12.4\\
        \hline
        venue\_area & Geographical key of the venue (combination of municipality name and H3 spatial index) & 13.9\\
        \hline
        venue\_name & Name of the venue & 18.3\\
        \hline
        venue\_desc & Short description of pictures in FoodX-251 randomly allocated to the venue & 686.7\\
        \hline
    \end{tabularx}
\end{table}

\subsubsection{Inference}
\label{inference}
As a preparation, we create a list of embedded values of all the POIs using the model trained in \ref{training}. Then, we take 3 steps for each sequence in the test dataset to predict the next POI: First, we keep the last POI of the sequence for performance evaluation (\ref{performance-evaluation}) and use the rest of the sequence. Then, we put this part of the sequence (except the last POI) into the trained model and get an embedded value of the next POI (prediction of the last POI). Finally, we select the most similar POI to this value from the list of embedded POIs in terms of cosine similarity.

\subsubsection{Performance evaluation}
\label{performance-evaluation}
For each sequence, we take the last POI as the true value and the POI selected in \ref{inference} as the predicted value. We select 4 typical metrics in recommendation tasks: nDCG, Recall, MRR, AUC. For nDCG and Recall, we calculate 2 different sample sizes at 10 and 50 for comparison.

\subsection{Performance Comparison}
Table \ref{table:performances} shows the model performances with and without POI picture descriptions against the test dataset. The overall performances are lower than the previous work with a deep learning model using the original Foursquare's Tokyo dataset where Recall@10 is 0.2172 \cite{islam2020survey, feng2020hme}, but we assume the performance difference is because of the differences of the venue categories and the user population (\ref{meta-data-and-interaction-data}). The model trained with picture descriptions significantly outperforms the model without them. nDCG@50 and Recall@50 are almost triple. The model without picture descriptions has no effectiveness to predict the next venue within 10 venues and its nDCG@10 and Recall@10 are both 0.

\begin{table}[h!]
    \caption{Model performances}
    \label{table:performances}
    \begin{tabularx}{\linewidth}{lXXXXXX}
        \hline
        \textbf{Picture desc.} & \multicolumn{2}{c}{\textbf{nDCG}} & \multicolumn{2}{c}{\textbf{Recall}} & \textbf{MRR} & \textbf{AUC}\\
        & @10 & @50 & @10 & @50 & & \\
        \hline
        With & 0.0019 & 0.0032 & 0.0049 & 0.0123 & 0.0017 & 0.545\\
        Without & 0 & 0.0011 & 0 & 0.0048 & 0.0009 & 0.520\\
        \hline
    \end{tabularx}
\end{table}

\section{Conclusion and Future Work}
In this paper, we test our novel concept of multimodal sequential recommendation of POI that can effectively learn textual attributes of venues and descriptions from visual information under geographical constraints. The model trained with picture descriptions has much higher metrics and this shows that our semi-multimodal model reflects actual human behaviours to choose a venue for a meal and that our path to a multimodal POI recommendation model is in the right direction. As the next steps, we will test with check-ins at other categories and combinations of them, and build a new architecture based on the previous works and our experiments. We will also add temporal information such as time span between check-ins and time frame of check-ins to the model to capture user behaviours more accurately on aspects of time and space which is a key feature of the mobility sector.

\begin{acks}
This work is partially supported by "Joint Usage/Research Center for Interdisciplinary Large-scale Information Infrastructures (JHPCN)" in Japan (Project IDs: jh231004 and jh241004). In addition, this work was partially supported by JSPS KAKENHI Grant Numbers JP21K17749 and 23K28098.
\end{acks}

\bibliographystyle{ACM-Reference-Format}
\bibliography{poi-recsys}


\appendix

\section{Food-related categories in the Foursquare dataset}
\label{appendix-food-categories}
We select the following 80 food-related categories in the Foursquare dataset (in the decending order of check-in frequency):
Ramen /  Noodle House, Japanese Restaurant, Food \& Drink Shop, Coffee Shop, Caf\'e, Fast Food Restaurant, Bar, Chinese Restaurant, Italian Restaurant, Restaurant, Indian Restaurant, Diner, BBQ Joint, Sushi Restaurant, Burger Joint, Bakery, Deli / Bodega, Asian Restaurant, Dessert Shop, Dumpling Restaurant, Steakhouse, Korean Restaurant, Sandwich Place, Donut Shop, Pizza Place, French Restaurant, Thai Restaurant, American Restaurant, Fried Chicken Joint, Seafood Restaurant, Candy Store, Beer Garden, Food, Tea Room, Soup Place, Ice Cream Shop, Spanish Restaurant, Snack Place, Mexican Restaurant, German Restaurant, Food Truck, Gastropub, Hot Dog Joint, Vietnamese Restaurant, Vegetarian / Vegan Restaurant, Breakfast Spot, Dim Sum Restaurant, Brazilian Restaurant, Middle Eastern Restaurant, Caribbean Restaurant, Tapas Restaurant, Cupcake Shop, Mediterranean Restaurant, Bagel Shop, Australian Restaurant, Eastern European Restaurant, Turkish Restaurant, Salad Place, Cajun / Creole Restaurant, Scandinavian Restaurant, Taco Place, Fish \& Chips Shop, Malaysian Restaurant, Latin American Restaurant, Portuguese Restaurant, South American Restaurant, African Restaurant, Burrito Place, Cuban Restaurant, Peruvian Restaurant, Ethiopian Restaurant, Moroccan Restaurant, Mac \& Cheese Joint, Swiss Restaurant, Argentinian Restaurant, Falafel Restaurant, Gluten-free Restaurant, Arepa Restaurant, Southern / Soul Food Restaurant, Afghan Restaurant

\section{Mappings between Foursquare and FoodX-251}
\label{appendix-4sq-foodx-mapping}
\begin{table}[ht]
    \caption{Mappings between Foursquare categories and FoodX-251 classes (in the alphabetical order)}
    \label{table:4sq-foodx-mapping}
    \begin{tabularx}{\linewidth}{lX}
        \hline
        \textbf{Foursquare categories} & \textbf{FoodX-251 classes}\\
        \hline
        'American Restaurant' & 'buffalo\_wing', 'clam\_chowder', 'fried\_egg', 'hamburger', 'hot\_dog', 'macaroni\_and\_cheese', 'salisbury\_steak', 'tetrazzini'\\
        \hline
        'Asian Restaurant' & 'pad\_thai', 'pho', 'spring\_roll', 'fried\_rice'\\
        \hline
        'BBQ Joint' & 'barbecued\_spareribs', 'barbecued\_wing'\\
        \hline
        'Bakery' & 'sausage\_roll'\\
        \hline
        'Breakfast Spot' & 'bacon\_and\_eggs', 'boiled\_egg', 'eggs\_benedict', 'ham\_and\_eggs'\\
        \hline
        'Burger Joint' & 'hamburger'\\
        \hline
        'Chinese Restaurant' & 'egg\_roll', 'fried\_rice', 'gyoza', 'moo\_goo\_gai\_pan', 'peking\_duck', 'spring\_roll', 'wonton'\\
        \hline
        'Dessert Shop' & 'apple\_pie', 'cheesecake', 'chiffon\_cake', 'chocolate\_cake', 'cupcake', 'fruitcake', 'macaron'\\
        \hline
        'Donut Shop' & 'donut'\\
        \hline
        'Dumpling Restaurant' & 'dumpling'\\
        \hline
        'Fast Food Restaurant' & 'fish\_stick', 'french\_fries', 'bacon\_lettuce\_tomato\_sandwich', 'club\_sandwich', 'grilled\_cheese\_sandwich', 'ham\_sandwich', 'hamburger', 'hot\_dog', 'lobster\_roll\_sandwich', 'pulled\_pork\_sandwich', 'victoria\_sandwich'\\
         \hline
    \end{tabularx}
\end{table}
\begin{table}[b]
    \begin{tabularx}{\linewidth}{lX}      
        \hline
        \textbf{Foursquare categories} & \textbf{FoodX-251 classes}\\
        \hline
         'French Restaurant' & 'beef\_bourguignonne', 'casserole', 'chicken\_cordon\_bleu', 'coq\_au\_vin', 'coquilles\_saint\_jacques', 'escargot', 'filet\_mignon', 'foie\_gras', 'lobster\_bisque', 'steak\_au\_poivre', 'steak\_tartare', 'veal\_cordon\_bleu', 'vol\_au\_vent'\\
        \hline
        'German Restaurant' & 'sauerbraten', 'sauerkraut', 'schnitzel'\\
        \hline
        'Hot Dog Joint' & 'hot\_dog'\\
        \hline
        'Ice Cream Shop' & 'ice\_cream'\\
        \hline
        'Indian Restaurant' & 'biryani', 'chicken\_curry'\\
        \hline
        'Italian Restaurant' & 'beef\_carpaccio', 'bruschetta', 'caprese\_salad', 'fettuccine', 'frittata', 'gnocchi', 'lasagna', 'linguine', 'panna\_cotta', 'penne', 'pizza', 'ravioli', 'rigatoni', 'risotto', 'spaghetti\_bolognese', 'spaghetti\_carbonara', 'tagliatelle', 'tiramisu', 'tortellini', 'ziti'\\
        \hline
        'Japanese Restaurant' & 'edamame', 'miso\_soup', 'sashimi', 'sukiyaki', 'takoyaki', 'tempura'\\    
        \hline
        'Korean Restaurant' & 'bibimbap'\\
        \hline
        'Mexican Restaurant' & 'chili', 'enchilada', 'guacamole', 'nacho', 'taco', 'tostada'\\
        \hline
        'Pizza Place' & 'pizza'\\
        \hline
        'Ramen / Noodle House' & 'ramen'\\
        \hline
        'Sandwich Place' & 'bacon\_lettuce\_tomato\_sandwich', 'club\_sandwich', 'grilled\_cheese\_sandwich', 'ham\_sandwich', 'lobster\_roll\_sandwich', 'pulled\_pork\_sandwich', 'victoria\_sandwich'\\
        \hline
        'Seafood Restaurant' & 'clam\_food', 'cockle\_food', 'crab\_food', 'lobster\_food', 'seaweed\_salad', 'sashimi'\\
        \hline
        'Spanish Restaurant' & 'adobo', 'paella'\\
        \hline
        'Steakhouse' & 'pepper\_steak'\\
        \hline
        'Sushi Restaurant' & 'sushi'\\
        \hline
        'Thai Restaurant' & 'pad\_thai', 'fried\_rice'\\
        \hline
        'Vegetarian / Vegan Restaurant' & 'beet\_salad', 'caesar\_salad', 'caprese\_salad'\\
        \hline
        'Vietnamese Restaurant' & 'pho'\\
        \hline
    \end{tabularx}
\end{table}









\end{document}